# Human Experts' Evaluation of Generative AI for Contextualizing STEAM Education in the Global South

Matthew Nyaaba[1], Macharious Nabang[2], Patrick Kyeremeh[3], Ibrahim Nantomah[4],
Collins Owusu-Fordjour[5], Martin Ako[6], Bismark Nyaaba Akanzire[7], Kassim Korah Nantomah[7]
Cecilia Issaka[4], Xiaoming Zhai[1]

[1] University of Georgia, USA
[2] Bagabaga College of Education, Ghana
[3] St. Joseph's College of Education, Ghana
[4] University for Development Studies, Ghana
[5] University of Education, Winneba, Ghana
[6] McGill University, Canada
[7] Gambaga College of Education, Ghana

**Abstract**

One of the key issues with STEAM education in many countries across the Global South is the abstract nature of the curriculum and the lack of contextualization of STEAM topics to meet sociocultural needs. This study investigates how human experts evaluate the capacity of Generative AI (GenAI) to contextualize STEAM education in the Global South. Grounded in human-centered and culturally responsive pedagogy, the study employed a convergent mixed-methods design involving four STEAM education experts who evaluated standardized Ghana National Council for Curriculum and Assessment (NaCCA) lesson plans alongside GenAI-generated lessons produced through a customized Culturally Responsive Lesson Planner (CRLP) powered by Interactive Semi-Automated prompts. The quantitative data were collected using a validated 25-item Culturally Responsive Pedagogy Rubric assessing bias awareness, cultural representation, contextual relevance, linguistic responsiveness, and teacher agency, while qualitative reflections provided interpretive depth on the pedagogical, cultural, and linguistic dynamics embedded in each lesson plan. Findings indicate that GenAI, particularly the CRLP tool, can meaningfully support STEAM instruction by connecting abstract curriculum standards to learners' sociocultural worlds. *Teacher Agency* emerged as the strongest domain, whereas *Cultural Representation* scored lowest. Across disciplines, CRLP-generated lessons were rated as more culturally grounded, student-centered, and pedagogically engaging than NaCCA lesson plans, integrated indigenous knowledge, bilingual elements, community artifacts, and locally grounded analogies, addressing long-standing concerns about the Westernized and decontextualized nature of STEAM education in Ghana. However, GenAI struggled to authentically capture Ghana's cultural pluralism, often producing surface-level references to local languages, histories, and identities. These weaknesses were most pronounced in Mathematics and Computing, where cultural nuance and diverse identity representation were least evident. Experts emphasized the need for continuous teacher mediation, community collaboration, and culturally attuned refinement of AI outputs. Future research should include classroom enactments, broaden expert participation, and explore model fine-tuning using Indigenous corpora to enhance cultural fidelity and linguistic precision in African educational contexts.

## Introduction

Contextualization in STEAM (Science, Technology, Engineering, Arts, and Mathematics) education has become an urgent pedagogical priority, particularly within the Global South, where educational content often fails to reflect learners' sociocultural realities (Ngman-Wara (Ngman-Wara, 2015; Nsengimana et al., 2020). In Ghana, for instance, Ngman-Wara (2015) revealed that junior high school science teachers had limited knowledge of contextualized science instruction, despite its alignment with national educational reforms. The study highlighted a significant disconnect between curriculum standards and actual classroom practice, where science was often taught devoid of local examples, analogies, or linguistic relevance (Ngman-Wara, 2015). This finding underscores a broader challenge across the Global South: while policy documents advocate for culturally and locally responsive pedagogy, teachers frequently lack the training, resources, and tools to implement it effectively, especially within STEAM domains where globalized content dominates (Nsengimana et al., 2020; Thibeault-Orsi, 2022).

The intensifying workload of teachers compounds this challenge (Do Minh et al., 2021). Lesson planning, a critical yet time-intensive task, often competes with grading, administration, and extracurricular duties (Sakamoto et al., 2024; Şimşek, 2025; Zheng & Stewart, 2024). Consequently, educators are increasingly turning to Generative AI (GenAI) tools to streamline lesson development. Studies such as van den Berg and du Plessis (2023) and Kerr and Kim (2025) show that GenAI can support lesson structure, resource generation, and language editing. However, these same studies caution that AI-generated lessons often lack context sensitivity, frequently omitting cultural and linguistic features essential to equitable STEAM education. As Hamouda et al. (2025) observed in their cross-African computing education review, efforts to localize content remain scattered and under-theorized, revealing the systemic difficulty of contextualizing STEAM teaching across diverse African classrooms.

One recent intervention addressing this gap is the Culturally Responsive Lesson Planner (CRLP), developed by Nyaaba and Zhai (2025). Grounded in culturally responsive pedagogy (CRP), their semi-interactive GenAI tool prompts teachers to embed local language, practices, and examples directly into AI-generated lesson plans(Nyaaba & Zhai, 2025; Nyaaba et al., 2024). Their findings showed that CRLP outputs outperformed generic GPT outputs in curriculum alignment, cultural relevance, and language use. However, their study stops short of comparing these CRLPs directly with human-created or standards-based lesson plans, a gap this study seeks to address. In this study, we seek to advance this line of inquiry by conducting a comparative analysis between GenAI-generated and curriculum-based (human-developed) lesson plans for STEAM education in Ghana. We employed human experts to generate STEAM lesson plans using Nyaaba and Zhai (2025) CRLP, evaluated them, and reflected on both the CRLP-generated lesson plans and the National Council for Curriculum and Assessment (NaCCA) lesson plans to determine their relative effectiveness in promoting culturally grounded and pedagogically sound instruction in STEAM education. Through this process, the study aims to uncover how GenAI-generated lessons can complement or enhance existing curricular materials and contribute to the advancement of equitable and contextually relevant STEAM education in the Global South. We emphasized human pedagogical judgment to ask the following questions.

1. How do experts evaluate the cultural responsiveness and pedagogical quality of GenAI-generated lesson plans?
2. In what ways do experts compare GenAI-generated lesson plans to standards-based (NaCCA) lesson plans?

**Literature Review**

**Theoretical Framework**

This study is grounded in two complementary theoretical perspectives: Culturally Responsive Pedagogy (CRP) and Human-Centered Artificial Intelligence (HCAI), situated within the broader concept of Human–AI Collaboration (Gay, 2015; Gay & Howard, 2000; Güvel et al., 2025). These frameworks provide the foundation for understanding how GenAI-generated lesson plans can align with local cultural and pedagogical contexts and how expert reflection can guide the responsible use of AI in education. This study adapts the AI and Culturally Responsive Assessment Framework developed by Nyaaba et al. (2024) as its theoretical foundation.

The framework integrates the principles of Culturally Responsive Pedagogy (CRP) with the evolving capabilities of GenAI to explore how GenAI can meaningfully support culturally grounded assessment. Their Nyaaba et al. (2024) framework embeds the classical works of Ladson-Billings (1995) and Gay (2018) to situate culture as a dynamic force that shapes both teaching and learning while matching the potentials of GenAI. The framework rests on five cultural tenets: *Indigenous language, Indigenous knowledge, ethnicity/race, religious beliefs, and community and family*, which serve as evaluative dimensions for determining the cultural responsiveness of both GenAI-generated and curriculum-based lesson plans (Nyaaba & Zhai, 2025; Nyaaba et al., 2024). These tenets ensure that instructional content represents diverse ways of knowing, respects linguistic and cultural plurality, and aligns with the lived realities of learners in Ghana and similar Global South contexts.

The study also draws from Human-Centered Artificial Intelligence (HCAI) to frame how teacher agency mediates the relationship between technology and pedagogy (Alfredo et al., 2024; Shneiderman, 2022). HCAI promotes AI systems that augment rather than replace human intelligence, preserving the educator's central role in judgment, creativity, and ethical decision-making, which emphasizes co-creation between human and machine intelligence (Zhang et al., 2024). In this study, experts act as reflective collaborators who assess and interpret GenAI-generated lesson plans relative to curriculum-based materials (Nyaaba & Zhai, 2025; Shneiderman, 2022).

**Possibilities and Challenges of AI for Lesson Planning**

Several recent studies underscore both the promise and limitations of GenAI in lesson planning, particularly in relation to cultural responsiveness, pedagogical soundness, and the need for human oversight (Nikolovski et al., 2025). Emerging scholarship has shown growing interest in the integration of GenAI into instructional planning, particularly in its capacity to support teachers through the automation of lesson design processes. Studies by van den Berg and du Plessis (2023) and Kerr and Kim (2025) highlight how tools like ChatGPT enable teachers, especially pre-service and overburdened educators, to generate lesson plans with greater efficiency. These tools assist in structuring content, suggesting topics, and formatting instructional sequences. However, both studies caution that such outputs often lack nuanced pedagogical reasoning or localized responsiveness, leading to concerns over the cultural appropriateness and educational quality of AI-generated materials (Kerr & Kim, 2025). Similarly, Sakamoto et al. (2024) report that while language differences may slightly influence content, a more pressing issue is GenAI's inability to represent learner diversity or reflect sociocultural context. Without targeted prompts or embedded values, AI-generated lessons tend to reproduce generic formats devoid of cultural grounding or ethical sensitivity, issues that are especially critical in multilingual and multicultural education settings (Choudhary, 2024).

To address these gaps, Nyaaba and Zhai (2025) propose a more advanced model in their design of a CRLP, an interactive GenAI tool grounded in Culturally Responsive Pedagogy (CRP). Their semi-automated prompting approach encourages teachers to input local linguistic, cultural, and curricular details before the lesson is generated. In their comparative study, CRLP was used to develop a Grade 7 Science lesson in Ghana's Ashanti Region (Nyaaba & Zhai, 2025; Nyaaba et al., 2024). The plan generated via CRLP was found to outperform a standard GPT-generated plan across key evaluative domains: cultural relevance, factual accuracy, and curriculum alignment. While this approach shows promise, their study focused on a single subject and grade level with limited human expert review and perspective. They recommended a comparative study between human-generated lesson plans and their CRLP tool. This study extends that recommendation by employing an in-depth human expert review alongside a comparison with a standard lesson plan (Nyaaba & Zhai, 2025).

In doing so, it makes a case for co-development models where AI serves as a flexible assistant rather than a prescriptive solution. It affirms that the most contextually relevant, culturally sensitive, and pedagogically sound lesson plans are not generated solely by machines but emerge through dynamic human-AI collaboration, where educators mediate and enrich AI outputs with local expertise, ethics, and lived experience (van den Berg & du Plessis, 2023; Kerr & Kim, 2025). UNESCO's Recommendation on the Ethics of AI opens by asserting that AI must serve human capabilities to foster inclusive, just, and sustainable futures, emphasizing a human-centred approach rooted in human rights, dignity, transparency, and accountability, principles that are operationalized across sectors like education, culture, labour, and health, and reinforced for teachers through guidelines to keep humans in the loop, uphold accountability, and safeguard learner rights and diversity (UNESCO, 2024).

**Challenges of STEAM Curriculum in the Global South**

STEAM education in the Global South remains fraught with contextual and structural challenges that limit its effectiveness and inclusivity (Gyamerah, 2025). Most STEAM curricula are modeled after Western epistemologies and pedagogical traditions, resulting in content and instructional approaches that often overlook indigenous knowledge, linguistic diversity, and community-based learning systems (Gyamerah, 2025). For example, Gyamerah (2024) found that both mathematics and science curricula emphasized the need for students to achieve global and Western values to become successful in their education. However, these curricula often overlook Ghanaian students' unique cultural values, competencies, and lived experiences, which do not conform to the prescribed global competencies (Anamuah-Mensah & Asabere-Ameyaw, 2006; Gyamerah, 2024). This omission risks reinforcing Western cultural norms as superior while devaluing Indigenous knowledge systems and competencies in STEAM education. Western-centric STEAM curriculum models often fail to resonate with local contexts, creating a disconnect between classroom learning and community realities (Gyamerah, 2024; Lafifa et al., 2023). Researchers advocate for culturally responsive pedagogy and the inclusion of Indigenous knowledge systems to make STEAM meaningful for African learners (Anamuah-Mensah & Asabere-Ameyaw, 2006; Gyamerah, 2024; Ogunniyi, 2023).

STEAM education also faces structural barriers due to rigid, siloed curricula that prioritize traditional subject separation and standardized assessments. For example, most African education systems still emphasize rote learning, leaving little room for interdisciplinary, project-based approaches central to STEAM (Ngman-Wara, 2015). Persistent issues such as under-resourced classrooms, limited access to digital technologies, and inadequate teacher training impede the development of inquiry-driven and cross-disciplinary learning (Anamuah-Mensah & Asabere-Ameyaw, 2006). Even when national curricula integrate STEAM principles, many educators struggle to contextualize them due to insufficient professional development opportunities and a lack of culturally relevant instructional materials (Ngman-Wara, 2015). Language of instruction, cultural diversity, and systemic inequities further complicate the delivery of equitable STEAM education across the Global South. The dominance of colonial languages as the medium of instruction marginalizes students' indigenous languages and cultural expressions, limiting comprehension and discouraging active participation (Gyamerah, 2024).

**GenAI's role in enhancing learning experiences in STEAM**

The swift progress in AI has revolutionized multiple sectors, including STEAM education (Mohana et al., 2022; Zhou, 2023). STEAM education offers various advantages, including the enhancement of critical thinking, creativity, intuition, emotion, and artistry while augmenting students' comprehension, interest, and potential in science and technology, thereby establishing a foundation for future innovators and entrepreneurs (Mohana et al., 2022). Consequently, educators and researchers have made good progress in applying AI within STEAM education (Nyaaba et al., 2024). In their study, Jang et al. (2022) emphasized the importance of AI content and methodological effectiveness in STEM education. Al-Zahrani et al. (2024b) conducted a study that delineated the trends in research regarding AI in STEAM education, utilizing the Web of Science database. Their review, encompassing 16 publications, demonstrated that AI technologies foster the development of cognitive skills, including computational and analytical thinking, bolster self-confidence, enhance pleasure and happiness among students, and further enrich their understanding of STEAM ideas (Al-Zahrani et al., 2024a). These conclusions hold significant significance for educators, practitioners, and policymakers in making informed judgments about the appropriate incorporation of AI in STEAM education.

Moreover, Lee and Zhai (2024) underscored the potential of incorporating GenAI models into STEAM instruction. In a comparable study, Lee et al. (2024) examined the application of GenAI in STEAM instruction for primary pupils. The study indicated that students perceived using AI for creative expression as fun and individualized, which may enhance the creative process. Their research suggests that GenAI can improve art teaching for primary school students (Lee et al., 2025). Price and Grover (2025) research brief provided significant insight into the possible influence of GenAI in aiding educators and enhancing STEM education. They emphasized that GenAI tools could assist educators in managing the intricacies of STEM instruction and enhance their ability to provide high-quality learning experiences tailored to the needs of all students in STEM classes.

Additionally, Price and Grover (2025) recognized the potential to enhance, expand, and unlock the beneficial consequences of GenAI on STEM education. However, they warned of the unequal distribution of advantages within the educational ecosystem. The authors suggest that historical precedence implies that those schools and STEM classrooms most likely to gain from GenAI help may experience the least improvement in student results. Gattupalli et al. (2025) incorporated GenAI technology into their technical report to create mathematical modules for students, offering clear, accessible, and engaging mathematical information. Al-Zahrani et al. (2024b) also conducted a systematic review of trends in AI utilization within STEAM, revealing that AI technologies foster the development of cognitive skills, including computational and analytical thinking, bolster self-confidence, elevate student satisfaction and enjoyment, and enrich comprehension of STEAM concepts.

**Method**

The study employed a convergent mixed-methods approach, which combined both quantitative and qualitative techniques to evaluate and interpret the effectiveness of CRLP-generated lesson plans compared with NaCCA-based lessons in STEAM subjects. This method was chosen because culturally responsive pedagogy involves both measurable instructional indicators and contextual, interpretive understanding (Creswell & Inoue, 2025). The convergence of numerical and narrative evidence ensured a comprehensive analysis, making this approach particularly relevant for examining how GenAI tools perform within Ghana's multilingual and culturally diverse educational context.

### Participant

We engaged four STEAM education experts who were purposively selected to bring together disciplinary mastery and deep contextual understanding of Ghanaian educational settings. Each participant had accumulated over five years of professional experience within their respective STEAM fields and possessed demonstrable cultural competence, reflected in their familiarity with local languages, schooling norms, and classroom practices. They also held substantive experience in teacher education, ranging from curriculum design to field experience evaluation, positioning them to critically interpret lesson plans through both pedagogical and culturally situated perspectives.

### Materials

Two major materials were used in this study. First, each reviewer independently selected Ghana's NaCCA lesson plan, which is standardized to follow national instructional guidelines. The selected lesson plan included, Mathematics (Basic 4 – *B4.1.2.6.1: Translate and solve word problems involving the four basic operations on whole numbers*), Science (Basic 4 – *B4.2.1.1.1: Demonstrate understanding of cyclic movements in the environment*), Creative Arts and Design (Junior High School – *Design in Nature and the Man-Made Environment*), and Computing (Junior High School – *Principles of Information Security: Confidentiality, Integrity, Availability*).

In the second phase, the same experts generated a corresponding lesson plan using the *Customized Culturally Responsive Lesson Planner GPT* embedded with *Interactive Semi-Automated (ISA) prompts*. These lesson plans were self-prompted by the reviewers to reflect culturally relevant themes, languages (e.g., Dagbani, Dagaare, and Akan), and community-based symbols, ensuring that GenAI outputs were pedagogically structured and locally grounded. In the third phase, we copied and pasted the GenAI-generated CRLP-generated lesson into editable documents together with their corresponding NaCCA lesson plans for comparative analysis (Appendix A & B).  In the fourth phase the reviewers evaluated lesson plans using a scoring rubric across a five-domain rubric (bias awareness, cultural representation, contextual relevance, linguistic responsiveness, and teacher agency) and wrote reflective narratives comparing both versions.

### Data Collection

To answer research question 1, a validated Culturally Responsive Pedagogy Rubric was used for quantitative data collection. The rubric consisted of 25 items, organized into five core domains: (1) *Bias Awareness* (e.g., avoiding stereotypes, representing diverse perspectives); (2) *Cultural Representation* (e.g., affirming local identities, challenging monocultural norms); (3) *Contextual Relevance* (e.g., reflecting local realities, using indigenous analogies); (4) *Linguistic Responsiveness* (e.g., bilingual support, accurate translation); and (5) *Teacher Agency* (e.g., adaptability, co-design with learners). Each item was scored using a 4-point ordinal scale, where 0 = Not Evident, 1 = Minimally Evident, 2 = Evident, and 3 = Strongly Evident. To answer research question 2 and to complement the rubric, and capture interpretive depth, a qualitative reflection sheet was provided for each expert. This allowed reviewers to justify their scores, offer narrative insights, identify potential cultural gaps or pedagogical limitations, and compare the instructional quality and responsiveness of the CRLP and NaCCA lesson formats.

**Data Analysis**

To answer Research Question 1, we conducted a structured quantitative analysis of the rubric-based evaluations provided by four STEAM experts across five culturally responsive domains: bias awareness, cultural representation, contextual relevance, linguistic responsiveness, and teacher agency. Mean scores were calculated at three levels: individual item averages across all subjects, subject-specific averages within each domain, and domain-wide averages across STEAM (mean of the individual score averages). These quantitative insights directly informed Research Question 2, which explored the themes emerging from expert reflections. Using Braun and Clarke's (2006) inductive thematic analysis, we treated each expert's commentary as a case, preserving verbatim expressions and aligning codes. The expert reflections consistently followed the logic and structure of the lesson plans, allowing the themes to emerge organically in relation to the instructional design components. The gerund findings from the thematic analytic process were visualized in a table using icons (✓ = strength, ✗ = gap) with illustrative quotes.

**Results**

**Expert Ratings**

In answering *RQ 1*, the item-level analysis of the GenAI-generated culturally responsive lesson plans across four STEAM domains reveals a nuanced distribution of strengths and limitations (see Table 1). Notably, *Teacher Agency* emerged as the strongest domain (M = 3.6), with experts consistently affirming that CRLP lessons support creative instructional design, co-design, and customization across all subjects. Likewise, *Contextual Relevance* (M = 3.4) scored highly, underscoring the CRLP model's capacity to integrate local realities, traditional knowledge, and curriculum-aligned content. Experts particularly valued how lesson elements like community artifacts, localized analogies, and indigenous examples grounded learning in familiar contexts, enhancing engagement and practical applicability. Moderate performance was observed in both *Bias Awareness* (M = 3.05) and *Linguistic Responsiveness* (M = 3.05). While GenAI-generated lessons largely avoided overt stereotypes and demonstrated strong multilingual features (especially bilingual assessments), concerns remained regarding the superficial treatment of historical injustices, limited attention to cultural idioms, and the occasional inaccuracy in local language translation.

In contrast, *Cultural Representation* (M = 2.25) emerged as the most underdeveloped domain, with mathematics and computing evaluations particularly highlighting gaps in affirming diverse identities, representing multiple cultural viewpoints, and challenging monocultural norms. The data suggests that while GenAI offers an efficient and structurally coherent foundation for lesson design, it tends to privilege standardized templates over culturally embedded content—especially in STEM subjects where abstract concepts often lack cultural contextualization. These findings emphasize the critical need for intentional teacher mediation to correct implicit biases, enrich cultural narratives, and ensure the equitable inclusion of historically marginalized groups. Importantly, the results also affirm that GenAI provides meaningful opportunities for teacher agency. Educators are empowered to revise and localize AI-generated outputs and must take responsibility for the cultural accuracy, contextual relevance, and pedagogical integrity of the final lesson materials used in their STEAM classrooms.

**Table 1**

*Expert Ratings by Rubric Category Across Subjects*

| Rubric Domain | Rubric Item | Computing | Science | Creative Arts | Mathematics | Item Average |
|---|---|---|---|---|---|---|
| Bias Awareness | Avoids stereotypes | 3 | 4 | 3 | 1 | 2.75 |
| | Avoids deficit narratives | 3 | 3 | 4 | 4 | 3.5 |
| | Reflects multiple ways of knowing | 4 | 3 | 4 | 3 | 3.5 |
| | Free from tokenism | 4 | 3 | 3 | 4 | 3.5 |
| | Sensitive to historical injustice | 3 | 2 | 2 | 1 | 2.0 |
| | **Domain Mean** | **3.4** | **3.0** | **3.2** | **2.6** | **3.05** |
| Cultural Representation | Affirms diverse identities | 1 | 2 | 4 | 1 | 2.0 |
| | Integral cultural contributions | 4 | 3 | 3 | 1 | 2.75 |
| | Includes multiple cultural examples | 3 | 1 | 3 | 3 | 2.5 |
| | Accurate and respectful portrayal | 2 | 2 | 4 | 1 | 2.25 |
| | Challenges monocultural norms | 2 | 2 | 3 | 0 | 1.75 |
| | **Domain Mean** | **2.4** | **2.0** | **3.4** | **1.2** | **2.25** |
| Contextual Relevance | Reflects local realities | 4 | 2 | 4 | 3 | 3.25 |
| | Aligns with curriculum | 4 | 4 | 4 | 4 | 4.0 |
| | Addresses community knowledge | 3 | 3 | 4 | 3 | 3.25 |
| | Localized analogies/symbols | 4 | 3 | 4 | 2 | 3.25 |
| | Adaptable across contexts | 3 | 3 | 4 | 3 | 3.25 |
| | **Domain Mean** | **3.6** | **3.0** | **4.0** | **3.0** | **3.4** |
| Linguistic Responsiveness | Supports native languages | 4 | 3 | 3 | 3 | 3.25 |
| | Appropriate syntax | 2 | 3 | 3 | 3 | 2.75 |
| | Accessible language | 3 | 3 | 4 | 3 | 3.25 |
| | Cultural idioms/expressions | 4 | 2 | 2 | 1 | 2.25 |

|  | Translation/multilingual support | 4 | 3 | 4 | 4 | 3.75 |
|---|---|---|---|---|---|---|
|  | **Domain Mean** | **3.4** | **2.8** | **3.2** | **2.8** | **3.05** |
| Teacher Agency | Customization by teachers | 3 | 4 | 4 | 3 | 3.5 |
|  | Supports creative design | 4 | 4 | 4 | 4 | 4.0 |
|  | Enables critique and revision | 3 | 4 | 4 | 3 | 3.5 |
|  | Encourages co-design | 3 | 4 | 4 | 4 | 3.75 |
|  | Reinforces professionalism | 3 | 2 | 4 | 4 | 3.25 |
|  | **Domain Mean** | **3.2** | **3.6** | **4.0** | **3.6** | **3.6** |

**Comparative Reflections**

The analysis of expert reflections reveals that the structure of the lesson plan components, *Objectives, Pedagogy, Student Engagement, Assessment, Language Use, Use of Resources, Cross-Cutting Issues, and Implementation Feasibility,* served as a guiding framework for their evaluations (see Table 2). This alignment shaped how thematic insights emerged across components, particularly when considering the cultural, linguistic, and practical dimensions of curriculum implementation.

*Objectives: Localizing Purpose for Relevance*

One of the strongest distinctions observed between the NaCCA and CRLP lesson plans lies in the formulation of lesson objectives. The NaCCA lesson objective focuses on enabling learners to discuss and describe the key principles of information security, specifically confidentiality, integrity, and availability. It aligns strictly with the national curriculum standard B7.3.3.1.1, which requires learners to recognize data threats and understand basic means of protection (see Figure 1). The cognitive demand is mainly at the recall and understanding level, emphasizing conceptual knowledge rather than applied practice. Cultural and contextual references are not included. The CRLP culturally responsive lesson objective expands the NaCCA requirement by enabling learners to explain, apply, and demonstrate the principles of information security using real-life examples from their own communities (see Figure 2). It asks learners to connect confidentiality, integrity, and availability to familiar local practices such as MoMo transactions, school records, and social media use, and to apply Dagbani vocabulary to reinforce understanding.

Human experts noted that the objectives often lacked cultural specificity. One reviewer stated that "the learning indicators in the NaCCA lesson plan are more generic and not culturally or context specific." This suggests a surface-level alignment with national standards but a disconnect from the lived experiences of learners. In contrast, CRLPs demonstrated stronger alignment with local realities, offering SMART objectives that were both clear and culturally anchored. One reviewer praised this clarity, noting, "The GenAI-generated lesson plan presents clearer lesson objectives that are SMART compared to the NaCCA curriculum." Another added, "The ChatGPT lesson plan had more context-specific objectives because they were stated within the local context, and examples could be given within the locality or community." These statements underscore the importance of embedding lesson goals within socio-cultural contexts, allowing learners to meaningfully connect curriculum content to their immediate environment.

| Date: 22nd , | DAY: | Subject: Computing | |
|---|---|---|---|
| Duration: | | Strand: Communication Networks | |
| Class: B7 | Class Size: | Sub Strand: Information Security | |
| Content Standard: B7.3.3.1. Recognize data threats and means of protection | Indicator: B7.3.3.1.1 Discuss the key principles of information security | | Lesson: 1 of 2 |
| Performance Indicator: Learners can describe the principles of information security. | | Core Competencies: CI 6.3: DL5.1: | |
| Reference: Computing Curriculum P.g. 18-20 | | | |
| Activities For Learning & Assessment | | Resources | Progression |
| *Starter (5 mins)* Using questions and answers, revise the previous lesson with learners. Share performance indicators and introduce the lesson. | | Pictures and videos | Describing the principles of information security. |

*Figure 1: NaCCA Computing Lesson Objective for Recognize data threats and means of protection*

> **JHS 1 Computing – Culturally Responsive Lesson Plan (45 Minutes)**
> **Topic: Principles of Information Security**
> **Key Principles: Confidentiality, Integrity, Availability**
> **1. Lesson Objectives (By the end of this lesson, learners will be able to):**
> - Explain the three key principles of information security: Confidentiality, Integrity, and Availabilit
> - Identify real-life examples of how these principles affect daily life in their communities
> - Demonstrate basic ways to protect personal information using relatable Dagbani life situations
>
> **2. Core Competencies:**
> - Digital literacy
> - Critical thinking
> - Personal development
> - Communication and collaboration
>
> **3. Materials Needed:**
> - Pictures or photocopies of MoMo transactions
> - Sample report cards / school files
> - Phones / illustrations of social media apps
> - Flashcards with Dagbani vocabulary
> - Blackboard or whiteboard

*Figure 2: CRLP Computing Lesson Objective for Recognize data threats and means of protection*

*Pedagogy: Shifting from Abstraction to Cultural Anchoring*

NaCCA lesson plans were described as pedagogically sound but disconnected from learners' cultural contexts. When it comes to Science pedagogy, the NaCCA activities emphasize structured, teacher-led exploration (see Figure 3). Learners are guided through demonstrations, observations, and simple investigations, such as watching pendulum motion or examining the maize life cycle. This approach focuses on conceptual understanding through direct observation, with limited integration of community contexts or local cultural knowledge. The CRLP science activities, however, maintain the essential NaCCA structure but expand it by embedding culturally grounded pedagogy, community relevance, and local experiential knowledge (see Figure 6). The activities invite learners to identify cyclic events from everyday life, such as seasonal farming, harvest festivals, and weather patterns familiar in their communities. Demonstrations, like showing evaporation and condensation, are immediately linked to Ghanaian environmental realities, such as rainfall cycles impacting maize and yam farming.

As one reviewer observed, "Examples… given in a different context… does not reflect the culture of the people." This indicates a tendency to generalize instructional strategies, potentially reducing learner engagement and comprehension. The CRLPs, on the other hand, employed a more culturally grounded methodology, integrating real-life examples that reflect the learners' environment. Experts consistently highlighted how CRLPs encouraged interactive, learner-centered approaches. One remarked, "The CRLP lesson employs learner-centered methods, including group discussions and real-life observation tasks." The intentional use of cultural artifacts, such as "symmetrical patterns in Dagaare homes and the artistic value of calabash engravings," added another layer of cultural relevance. Such design choices promote identity-affirming instruction and deepen conceptual understanding by anchoring learning in familiar practices.

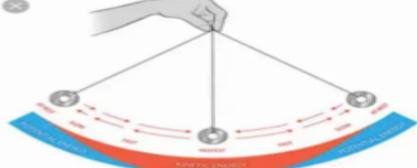

*Figure 3: NaCCA Science Activities on Classify materials into liquids, solids and gases*

> **Main Activities (40 minutes):**
> Activity 1: Identifying Cyclic Events (15 minutes)
> Group Work: Divide the class into small groups and ask them to list and describe any recurrent events they notice in their surroundings (e.g., seasonal farming cycles, traditional festivals linked to the harvest season).
>
> Class Discussion: Groups present their findings, and the teacher builds on their ideas to explain broader cyclic processes like day and night, seasons, and the water cycle.
>
> **Activity 2: Exploring the Water Cycle (15 minutes)**
> Demonstration: Use a bowl of water, a heat source (such as a candle or hot water), and a cool surface to show the process of evaporation and condensation, linking it to the water cycle in Ghana (e.g., explaining the importance of rainfall in agriculture, especially for crops like maize and yam).
>
> Link to Local Context: Discuss how rainfall patterns affect farming practices in Ghanaian communities.

*Figure 4: CRLP Science activities on Classify materials into liquids, solids and gases*

*Engagement and Contextualization: Making Learning Relevant and Active*

When it comes to student engagement, the NaCCA lesson adopts a traditional, structured approach centered on question-and-answer review, brainstorming, pair research, and teacher-guided discussions (see Figure 5). Engagement is primarily cognitive and academic, focusing on explanation and description, supported by general resources such as pictures and videos. However, the activities do not explicitly tap into learners' cultural experiences, home practices, or local communication norms. In contrast, the CRLP lesson expands engagement through deep cultural grounding, linguistic inclusion, and experiential relevance (see Figure 6). The introduction uses Dagbani-English mixed prompts, drawing learners into the lesson through relatable questions about MoMo PINs, privacy of family health records, and daily digital experiences in the community. The presentation phase embeds the CIA Triad into Dagbani concepts, local idioms, and community-based examples, enabling students to understand confidentiality, integrity, and availability through everyday scenarios such as school report cards, hospital records, and mobile money transactions. Learner activities require small groups to analyze localized scenarios, identify which security principle is violated, propose solutions, and act out the scenarios using Dagbani.

The CRLPs were consistently described as more engaging and context-sensitive than the NaCCA lesson plans. While NaCCA was acknowledged for having an "inquiry-based structure," it was also criticized for relying on "generalized examples." This lack of contextual detail weakened the lesson's ability to resonate with students. In contrast, reviewers lauded the CRLPs for being "activity-oriented and engaging" and for designing lessons that "acknowledge the lesson context." One expert elaborated, "Students are more likely to engage in meaningful learning given the GenAI-generated lesson plan's engaging and activity-oriented nature." This suggests that CRLPs more effectively bridge theoretical content and everyday experience, motivating students through culturally and socially relevant activities.

| Activities For Learning & Assessment | Resources | Progression |
|---|---|---|
| **Starter (5 mins)**<br><br>Using questions and answers, revise the previous lesson with learners.<br><br>Share performance indicators and introduce the lesson.<br><br>**Main (35 mins)**<br><br>Brainstorm learners to explain the meaning of information security.<br>Information security covers the tools and processes that organizations use to protect information.<br><br>Research in pairs the key principles of information security.<br>Example: confidentiality, integrity and availability<br><br>Guide learners to discuss the three key principles of information security.<br>Confidentiality<br>Confidentiality measures are designed to prevent unauthorized disclosure of information. The purpose of the confidentiality principle is to keep personal information private and to ensure that it is visible and accessible only to those individuals who own it or need it to perform their organizational functions. | Pictures and videos | Describing the principles of information security. |

*Figure 5: NaCCA Engagement Computing Activities on Recognize data threats*

4. Lesson Introduction (5 mins)
Warm-up discussion (Use Dagbani & English mix). Teacher prompts:
- "Zɔŋ mali ka o zaɣim bia bɔhi m-fɔ ka yel' ni bia mo MoMo PIN?"
*(Have you ever seen someone telling another person their MoMo PIN?)*
- "Fo yɛ ka ni yeli confidentiali?"
*(Do you know what it means for something to be confidential?)*

Relate it to real experiences in the community:
"Suppose a friend takes your phone and sees your family's health records or reads your messages—how would you feel?"

5. Lesson Presentation (20 mins)
Introduce the CIA Triad (use visual chart)

| Principle | Dagbani Meaning/Idea | Local Example (Dagbon) |
|---|---|---|
| Confidentiality | *Yam pahi, ti kpahi o daa* (Keep it private, don't say it) | MoMo PIN yel' bia ka bɔli |
| Integrity | *N ti mali suɣ'ni* (It has not been changed) | Report card ka n yoli nya |
| Availability | *Ka ni mali ka fo pahi la* (It's there when you need it) | Hospital data ka n ti pahi yɛli doctor yɛr'bu |

💬 Questions for class discussion:
- "Dɛn be wɔta MoMo PIN yel' ka confidentiali pahi?"
- "If a teacher changes your marks without telling you, does it still have integrity?"
- "What happens if your school report file goes missing — is it still available?"

6. Learner Activity (15 mins)
👥 Group Work (5 learners per group) – Give each group a simple scenario in Dagbani/English.
Scenario Examples:
1. A boy shares his school login with a friend. (Confidentiality issue)
2. Someone changes your grades on the school portal. (Integrity issue)
3. School attendance file gets locked in the headteacher's drawer. (Availability issue)

*Figure 6: CRLP Engagement Computing Activities on Recognize Data Threats*

*Language Use: Inclusion Through Localization, with Translation Caveats*

The NaCCA lesson uses English only as the medium of instruction. All explanations, definitions, and examples of information security are delivered entirely in formal English, with no incorporation of local languages or culturally familiar expressions (see Figure 5). This keeps the lesson aligned with national academic language standards but limits opportunities for learners to connect the concepts to the ways they speak and interact in their daily lives. The CRLP lesson, however, intentionally uses a bilingual approach, combining English and local languages such as Dagbani throughout the instructional process for computing principles, confidentiality, integrity, and availability are explained using Dagbani phrases, local idioms, and examples from everyday community interactions such as MoMo use, report cards, and hospital records. Learners engage in discussions, group work, and role-play using a Dagbani–English mix (see Figure 6).

Experts viewed this as a strength, positioning CRLPs as more linguistically responsive than the NaCCA curriculum, which "does not reflect any specific culture or context." However, concerns were raised about the accuracy and inclusiveness of translation. One reviewer pointed out, "Some learners are still excluded because they don't understand the local language," while another observed, "Most of the translations are not accurate." Despite these limitations, the inclusion of local expressions and examples was praised: "The inclusion of Dagaare examples and cultural symbols provides an entry point for inclusive engagement." Thus, while CRLPs hold potential for linguistic equity, effective implementation depends on teacher mediation and local adaptation.

*Resources: Sustainability and Cultural Continuity*

A key area where CRLPs outperformed NaCCA was in the selection and use of instructional materials. In the NaCCA lesson, resources are mainly generic and standardized, such as pictures of birds, leaves, trees, landscapes, textbook diagrams, and online images (e.g., Pinterest). These materials support understanding of design concepts but are not culturally specific and do not reflect Ghanaian or local artistic traditions (see Figure 7). In contrast, the CRLP lesson uses more local, culturally grounded materials, including pictures of familiar natural forms from the Upper West Region, traditional Dagaare weaving and pottery, calabashes with engravings, woven baskets, and natural materials like leaves, sticks, seeds, and clay (see Figure 8). These resources make learning contextual, identity-affirming, and practical, allowing learners to connect design concepts to their own community and heritage. While NaCCA often "required external teaching packs or imported resources," the CRLPs incorporated "resources… easily found within the locality and would not be expensive to get." Experts praised the CRLPs for promoting both sustainability and cultural continuity. One reviewer noted that "the resources… are highly accessible, drawn from the local environment and traditional material culture." Another elaborated, "By suggesting tools such as locally sourced leaves, sticks, and handmade artifacts, the lesson reinforces sustainability and cost-effectiveness." These reflections highlight the value of culturally embedded resource use, especially in low-resource settings.

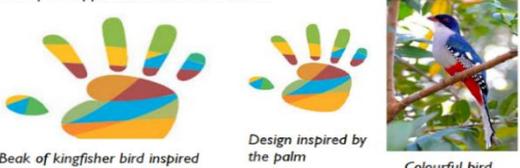

*Figure 7: NACCA Resources for Arts Education*

**Lesson Plan Generated with the Culturally Responsive Lesson Planner (CRLP)**
**School Level**: Junior High School
**Grade**: JHS
**Subject**: Creative Arts and Design
**Topic**: Design in Nature and Man-Made Environment
**Class Size**: 45 students
**Age Group**: 13-16 years
**Duration**: 60 minutes
**Location**: Daffiama-Bussie-Issa District, Upper West Region, Ghana

-------------------------------------------------------------------------------------

**General Objective:**
Students will demonstrate understanding of design as a concept in relation to the elements (dots, lines, and shapes) and principles (balance, rhythm, repetition) of design, and as a medium for creative expression of design in nature and the man-made environment.

**Specific Objectives:**
By the end of the lesson, students should be able to:

1. Identify and describe elements and principles of design.
2. Differentiate between natural and man-made designs using local examples.
3. Create a simple design inspired by nature or traditional Dagaare man-made art forms.

**Materials:**

- A4 sheets, pencils, markers, crayons
- Pictures of natural forms (leaves, animals)
- Images of traditional Dagaare weaving, pottery, and architecture
- Calabashes with engravings, woven baskets
- Locally sourced natural materials (leaves, sticks)

*Figure 8: CRLP Resources for Arts Education*

*Cross-Cutting Issues and Bias Awareness: The Need for Teacher Oversight*

The NaCCA lesson plans were seen as systematically addressing national cross-cutting priorities, like inclusion, sustainability, and digital literacy, CRLPs presented a more mixed picture. Reviewers acknowledged the innovative potential of GenAI but also cautioned that the CRLPs "seem not sensitive to the issue of stereotypes." Bias in content generation was a noted concern: "May inadvertently contain Eurocentric examples, gender stereotypes, or cultural omissions." As such, the CRLP approach shifts responsibility onto teachers as ethical evaluators. One reviewer emphasized, "Teachers must scrutinize every element… replacing generic or potentially biased examples with inclusive alternatives." These reflections point to a recurring theme: teacher agency is not optional but essential in mediating GenAI-generated content.

*Implementation Feasibility: Efficiency Meets Adaptability*

NaCCA lesson plans, while structured and policy-aligned, were sometimes described as difficult to adapt and overly reliant on infrastructure. CRLP's lesson plans, by contrast, were praised for enhancing implementation feasibility. One expert noted, "The CRLP lesson plan enhances practical usability and immediate classroom relevance." In particular, the use of GenAI was seen as a time-saving tool. As one reviewer commented, "Automation streamlines preparation and frees up valuable time." Yet, even this strength requires careful handling, as teachers must adapt the content to ensure contextual and cultural fit. Thus, CRLPs offer both instructional efficiency and pedagogical flexibility, provided educators are equipped to guide their application responsibly and ethically in their classroom.

**Table 2**

*Comparative Analysis of NaCCA and CRLP (GenAI-Generated) Lesson Plans with Expert Commentary*

| Theme/Component | NaCCA (Standards-Based) | CRLP (GenAI-Generated) | STEAM Relevance | Key Insight (with Supporting Quotes from the Experts) |
|---|---|---|---|---|
| Objectives | ✔ Curriculum-aligned but generic ✘ Lacks cultural specificity | ✔ SMART, clear, and rooted in local context ✔ Links objectives to community knowledge | Math, Science, Arts, Tech | *CRLP better connects content to student realities; NaCCA lacks contextual depth.* "The learning indicators in the NaCCA lesson plan are more generic and not culturally or context specific." "The ChatGPT lesson plan had more context-specific objectives because they were stated within the local context and examples could be given within the locality or community." |
| Pedagogy | ✔ Structured ✘ Uses generalized examples | ✔ Real-life, culturally grounded examples ✔ Group-based, learner-centered methods | Science, Arts, Math | *CRLP promotes identity-affirming learning; NaCCA lacks cultural responsiveness.* "Examples… given in a different context… does not reflect the culture of the people." "The CRLP lesson employs learner-centered methods, including group discussions and real-life observation tasks." |
| Engagement & Context | ✘ Limited contextualization ✔ Inquiry-based structure | ✔ Activity-oriented and engaging ✔ Acknowledges local realities | Math, Science, Arts, Tech | *CRLP fosters more meaningful learning and motivation.* "The GenAI-generated lesson plan appears to be more engaging and activity-oriented compared to the NaCCA curriculum." "The lesson seems to acknowledge the lesson context…" |
| Assessment | ✘ Generic, one-format assessment ✘ English-only | ✔ Formative, flexible, bilingual assessment ✔ Includes oral and peer formats | Math, Science, Arts, Tech | *CRLP supports equity in multilingual classrooms.* "The lesson gives opportunity for the learner to provide his/her response in either English or the Dagbani local language." "The assessment is done both in the local language and in English… This means that the assessment is both cultural and context responsive." |

| Category | NaCCA | CRLP (GenAI) | Subjects | Summary & Evidence |
|---|---|---|---|---|
| Language Use | ✘ English-only<br>✓ Standardization | ✔ Bilingual (local + English)<br>✘ Translation challenges<br>✘ Potential exclusion if not adapted | Math, Science, Arts, Tech | *CRLP is linguistically inclusive but needs accurate translation and balance.*<br>"Most of the translations are not accurate."<br>"Some learners are still excluded because they don't understand the local language."<br>"The inclusion of Dagaare examples and cultural symbols provides an entry point for inclusive engagement." |
| Resources | ✘ Sometimes requires imported tools | ✔ Uses local, sustainable, low-cost materials<br>✔ Aligns with traditional knowledge | Science, Arts | *CRLP better supports low-resource schools.*<br>"Resources were easily found within the locality and would not be expensive to get."<br>"By suggesting tools such as locally sourced leaves, sticks, and handmade artifacts, the lesson reinforces sustainability and cost-effectiveness." |
| Cross-Cutting Issues | ✔ Addresses inclusion, sustainability, digital literacy | ✘ Less emphasis on policy-based issues<br>✘ Possible stereotype bias in AI output | Math, Science, Arts, Tech | *NaCCA aligns with national priorities; CRLP needs teacher adaptation.*<br>"Deliberate efforts are made to address cross-cutting issues in the NaCCA lesson plan."<br>"Teachers need to adapt the GenAI-generated lesson by eliminating and addressing any issues of bias." |
| Implementation Feasibility | ✘ May require structured packs or infrastructure | ✔ Highly adaptable and efficient<br>✔ GenAI streamlines planning<br>✔ Saves teacher time | Math, Science, Arts, Tech | *CRLP enhances classroom readiness and practical planning.*<br>"The CRLP lesson plan enhances practical usability and immediate classroom relevance."<br>"Automation streamlines preparation and frees up valuable time." |

**NB:** *The icons indicators in the table reflect key evaluation patterns: strengths are marked in (✔), weaknesses or gaps in (✘)*

## Discussion

This study set out to examine the extent to which GenAI-produced lesson plans align with culturally responsive pedagogical principles across STEAM disciplines in Ghana, using a combination of expert ratings and comparative qualitative reflections. It responds directly to Nyaaba and Zhai's (2025) recommendation to compare GenAI-generated lessons with standardized human-developed plans and to foreground human evaluation in assessing GenAI's pedagogical usefulness. In doing so, the study adopts a perspective consistent with human-centered and value-sensitive design, which positions teachers' judgment, contextual insight, and intentional decision-making at the center of technological integration (Alfredo et al., 2024; Shneiderman, 2022). Culturally responsive pedagogy, as defined by Gay (2015), provides the interpretive framework for determining the cultural and contextual appropriateness of AI-generated lessons (Gay, 2025).

Moreover, the findings broadly suggest that GenAI has the potential to support STEAM teachers in Ghana, addressing long-standing challenges of the abstract and Western-centric STEAM education context (Gyamera, 2025). Furthermore, the concerns raised by Ngaman-Wara (2025) regarding the limited STEAM pedagogical knowledge among teachers in Ghana appear to have a practical solution through the capabilities of GenAI, particularly when used with the CRLP tool for lesson design. This integration positions GenAI as a meaningful avenue for reinforcing teacher preparation and adoption of GenAI in enhancing their instructional quality and bridging the persistent gaps in STEAM education across the country (Zhai, 2024). This finding reinforces the view that GenAI, when integrated thoughtfully, can serve as a pedagogical partner rather than a prescriptive authority (Nyaaba & Zhai, 2025; Roe & Perkins, 2024). Roe and Perkins (2024) similarly note that while GenAI has the potential to enhance instructional efficiency, its value is maximized only when human educators maintain control over judgment, creativity, and cultural interpretation.

Furthermore, GenAI's ability to integrate community artifacts, indigenous knowledge systems, and locally adapted analogies demonstrates that AI tools can meaningfully connect abstract curriculum standards to the everyday realities of Ghanaian learners. This directly addresses longstanding concerns about the STEAM curriculum being overly westernized, abstract, and disconnected from learners' cultural worlds (Lafifa et al., 2023; Anamuah-Mensah & Asabere-Ameyaw, 2006). This finding was consistent across all STEAM disciplines examined in the study, revealing a cross-disciplinary pattern similar to what Nyaaba and Zhai (2025) observed in their empirical application of the CRLP model in Ghana's Ashanti Region, where GenAI-generated lessons outperformed generic plans in relevance, accuracy, and curriculum fit.

Despite GenAI's capacity to support culturally responsive pedagogy, GenAI-produced lesson plans still struggled to affirm diverse identities or fully reflect the epistemologies of Ghana's plural societies. This aligns with Jayasinghe et al. (2025), who warn that without intentional safeguards against tokenism, GenAI systems risk reproducing dominant cultural logics. Although

CRLP outputs incorporated bilingual elements and native language references, they fell short when dealing with idiomatic expressions, complex historical narratives, and culturally embedded translations. Jeon et al. (2025), drawing on a translanguaging perspective, attribute these limitations to GenAI's reliance on dominant-language training data, which restricts its ability to interpret local nuance. Our findings support this critique and underscore the need for deliberate teacher mediation to ensure linguistic inclusion is substantive rather than symbolic. Likewise, Dennison et al. (2025) emphasize that culturally aligned AI systems in non-Western contexts require sustained collaboration with local educators and communities, reinforcing the position that AI alone cannot adequately represent the cultural multiplicities of the Global South.

In terms of pedagogical orientation, CRLPs were consistently described as more culturally grounded and student-centered than their NaCCA counterparts. The use of real-life examples, group discussion formats, and tangible cultural references like Dagaare symbols and calabash engravings added depth to abstract concepts. He et al. (2025) similarly observe that teacher-AI co-creation can result in learning materials that reflect community values and support deeper conceptual understanding. The assessment dimension of CRLPs was also praised, particularly for incorporating formative, bilingual, and multimodal strategies that valued student expression in both English and local languages. This stands in contrast to the more rigid, English-only approaches observed in NaCCA plans. Again, these results align with Wang et al. (2025) and Peterson and Jensen (2025), both of whom emphasize the importance of linguistic inclusion and teacher autonomy in assessment design. Despite these strengths, the limited treatment of national cross-cutting issues in CRLPs, such as gender equity, sustainability, and digital literacy, remains a concern. Unlike NaCCA plans, which addressed these priorities more systematically, GenAI - lesson plans depended heavily on the specificity of prompts and the vigilance of the teacher(Kerr & Kim, 2025; Nyaaba & Zhai, 2025).

**Conclusion**

This study set out to examine, with both numerical clarity and cultural sensitivity, how a customized GenAI Lesson Planner performs when placed side-by-side with Ghana's NaCCA lesson plans across four STEAM subjects. By engaging four experienced STEAM experts, each deeply grounded in Ghanaian educational realities and fluent in the cultural rhythms of our classrooms, the study brought together the strengths of quantitative scoring and the reflective depth of qualitative insights. Through rubric evaluations, thematic reflections, and direct comparison of paired lesson plans, we were able to understand not only *what* GenAI produces, but *how* it aligns, or struggles to align, with the lived experiences, languages, and learning traditions of West African learners.

The findings show clearly that GenAI-generated CRLPs hold significant promise. They offered sharper, culturally rooted objectives; richer contextualization grounded in familiar artifacts, local analogies, and community knowledge; and more flexible, bilingual assessments that speak to the multilingual reality of our classrooms from Tamale to Wa, from Kumasi to Accra. These strengths echo the aspirations of many African educators who desire tools that honour our knowledge systems while supporting innovative teaching practice. Yet the study also shows that GenAI is not without its blind spots. Cultural representation remained the least-developed domain, with notable gaps in affirming multiple identities and accurately capturing cultural nuances, especially in Mathematics and Computing. Furthermore, translation challenges, subtle stereotype risks, and occasional misalignment with local idioms remind us that AI, on its own, cannot fully understand the depth of our cultural story.

As with many aspects of life and learning in West Africa, balance becomes essential. GenAI offers speed, structure, and creative possibility, but teachers remain the custodians of context, language, and cultural truth. This study, therefore, recommends that GenAI be embraced as a *co-designer* rather than an unquestioned authority. Teacher education programs, especially in the Global South, should embed GenAI literacy that emphasizes ethical use, contextual judgement, and bias detection. Future work should expand the expert pool, include classroom enactment data, and explore fine-tuning models with Indigenous and community-based corpora to strengthen cultural fidelity. The study affirms that GenAI can meaningfully support culturally responsive teaching, but only when guided by the lived experience of teachers who understand the heartbeat of the communities they serve. The path forward lies in a thoughtful partnership between human insight and technological innovation, ensuring that AI becomes a tool for equity, cultural continuity, and educational transformation across Ghana and the wider African region.

**Limitations of the Study**

Even though this study offers valuable insight into the potential of GenAI-generated culturally responsive lesson plans within Ghana's STEAM classrooms, several limitations must be acknowledged. First, the sample size was intentionally small, involving four STEAM experts who brought strong cultural and disciplinary expertise but limited the breadth of perspectives. Their evaluations reflect deep contextual knowledge, yet a larger and more diverse expert pool, including classroom teachers, language specialists, and curriculum officers, would broaden the interpretive range and strengthen the generalizability of findings. Second, the study focused solely on lesson plans rather than classroom enactment. Although expert critiques provide a rich understanding of cultural alignment and pedagogical strength, future work should examine how CRLPs function in real instructional settings, how learners respond to them, and how teachers adapt them under ordinary classroom constraints.

Third, GenAI outputs were produced through prompt strategies rooted in Ghanaian languages and cultural cues, but the models themselves were not fine-tuned on local corpora. This limitation affected translation accuracy, depth of cultural representation, and the nuanced handling of Indigenous expressions, suggesting that the cultural fidelity of CRLPs is constrained by the underlying training data. Additionally, the study assessed only four STEAM subjects, leaving open questions about how GenAI performs in humanities, vocational subjects, and early childhood contexts where cultural knowledge is even more integrally tied to teaching and learning. Finally, the rubric employed—though validated, captures observable elements of cultural responsiveness but may not fully account for implicit cultural meanings, historical dynamics, or worldview representations that are essential in many African learning communities.

**References**

Al-Zahrani, A., Khalil, I., Awaji, B., & Mohsen, M. (2024a). AI technologies in STEAM education for students: Systematic literature review. *Journal of Ecohumanism*, *3*(4), 3380–3397. https://doi.org/10.62754/joe.v3i4.4041

Al-Zahrani, A., Khalil, I., Awaji, B., & Mohsen, M. (2024b). AI technologies in STEAM education for students: Systematic literature review. *Journal of Ecohumanism*, *3*(4), 3380–3394. https://doi.org/10.62754/joe.v3i4.4041

Alfredo, R., Echeverria, V., Jin, Y., Yan, L., Swiecki, Z., Gašević, D., & Martinez-Maldonado, R. (2024). Human-centred learning analytics and AI in education: A systematic literature review. *Computers and Education: Artificial Intelligence*, *6*, 100215. https://doi.org/https://doi.org/10.1016/j.caeai.2024.100215

Anamuah-Mensah, J., & Asabere-Ameyaw, A. (2006). The fusion of modern and indigenous science and technology: how should it be done? *African Journal of Educational Studies in Mathematics and Sciences*, *2*(1), 49–58. https://doi.org/10.4314/ajesms.v2i1.38587

Choudhary, T. (2024). Reducing Racial and Ethnic Bias in AI Models: A Comparative Analysis of ChatGPT and Google Bard.

Creswell, J. W. (2021). *A concise introduction to mixed methods research*. SAGE publications.

Creswell, J. W., & Inoue, M. (2025). A process for conducting mixed methods data analysis. *Journal of General and Family Medicine*, *26*(1), 4–11.

Dennison, D. V., Jain, M., Ganu, T., & Vashistha, A. (2025). *Designing culturally aligned AI systems for social good in non-Western contexts*. https://doi.org/10.48550/arXiv.2509.16158

Do Minh, H., Do Minh, H., Pham Trut, T., & Pham Trut, T. (2021). Reflective Teaching Perceived and Practiced by EFL Teachers - A Case in the South of Vietnam. *International Journal of Instruction*. https://doi.org/10.29333/iji.2021.14219a

Gattupalli, S., Maloy, R., & Edwards, S. (2025). *Fine-tuning a GenAI chatbot in the development of math modules for early elementary learners*. https://doi.org/10.7275/b72v-qp94

Gay, G. (2015). The what, why, and how of culturally responsive teaching: International mandates, challenges, and opportunities. *Multicultural Education Review*, *7*(3), 123–139.

Gay, G. (2018). *Culturally responsive teaching: Theory, research, and practice* (3 ed.). Teachers College Press.

Gay, G., & Howard, T. C. (2000). Multicultural teacher education for the 21st century. *The teacher educator*, *36*(1), 1–16.

Güvel, M., Kiyak, Y., Varan, H., Sezenöz, B., Coskun, Ö., & Uluoglu, C. (2025). Generative AI vs. human expertise: a comparative analysis of case-based rational pharmacotherapy question generation. *European Journal of Clinical Pharmacology*, *81*(6), 875–883. https://doi.org/10.1007/s00228-025-03838-2

Gyamerah, K. (2024). *The Role of African Indigenous Knowledge Systems and Pedagogies in Decolonizing and Transforming Mathematics and Science Learning in Ghana* (Publication Number 31366694) [Ph.D., Queen's University (Canada)]. ProQuest Dissertations & Theses A&I; ProQuest Dissertations & Theses Global. Canada -- Ontario, CA. https://www.proquest.com/dissertations-theses/role-african-indigenous-knowledge-systems/docview/3110362788/se-2?accountid=14537

https://galileo-uga.primo.exlibrisgroup.com/openurl/01GALI_UGA/01GALI_UGA:UGA?url_ver=Z39.88-2004&rft_val_fmt=info:ofi/fmt:kev:mtx:dissertation&genre=dissertations&sid=ProQ:ProQuest+Dissertations+%26+Theses+Global&atitle=&title=The+Role+of+African+Indigenous+Knowledge+Systems+and+Pedagogies+in+Decolonizing+and+Transforming+Mathematics+and+Science+Learning+in+Ghana&issn=&date=2024-01-01&volume=&issue=&spage=&au=Gyamerah%2C+Kenneth&isbn=9798384344193&jtitle=&btitle=&rft_id=info:eric/&rft_id=info:doi/

Gyamerah, K. (2025). Rethinking the potential role of African Indigenous knowledge systems in transforming STEM education. *Diaspora, Indigenous, and Minority Education*, 1–13. https://doi.org/10.1080/15595692.2025.2553599

Hamouda, S., Marshall, L., Sanders, K., Tshukudu, E., Adelakun-Adeyemo, O., Becker, B. A., Dodoo, E. R., Korsah, G. A., Luvhengo, S., Ola, O., Parkinson, J., & Sanusi, I. T. (2025). *Computing Education in African Countries: A Literature Review and Contextualised Learning Materials* 2024 Working Group Reports on Innovation and Technology in Computer Science Education, Milan, Italy. https://doi.org/10.1145/3689187.3709606

He, K., Liu, X., Xu, Y., Bustamante, A. S., & Warschauer, M. (2025). "Carlitos the Curious Caterpillar": Exploring Teacher-AI Co-Creation of Culturally Responsive Educational Materials for Young Learners. In *Proceedings of the 24th Interaction Design and Children* (pp. 236–254). Association for Computing Machinery. https://doi.org/10.1145/3713043.3727056

Jang, J., Jeon, J., & Jung, S. K. (2022). Development of STEM-based AI education program for sustainable improvement of elementary learners. *Sustainability*, *14*(22), 15178.

Jayasinghe, S., Arm, K., & Gamage, K. A. A. (2025). Designing Culturally Inclusive Case Studies with Generative AI: Strategies and Considerations. *Education Sciences*, *15*(6), 645.

Jeon, J., Wei, L., Tai, K. W. H., & Lee, S. (2025). Generative AI and its dilemmas: Exploring AI from a translanguaging perspective. *Applied Linguistics*, *46*(4), 709–717. https://doi.org/10.1093/applin/amaf049

Kerr, R. C., & Kim, H. (2025). From Prompts to Plans: A Case Study of Pre-Service EFL Teachers' Use of Generative AI for Lesson Planning. *English Teaching*, *80*(1), 95–118.

Ladson-Billings, G. (1995). Toward a theory of culturally relevant pedagogy. *American educational research journal*, *32*(3), 465–491.

Lafifa, F., Rosana, D., Suyanta, S., Nurohman, S., & Dwi Astuti, S. R. (2023). Integrated STEM Approach to Improve 21st Century Skills in Indonesia: A Systematic Review. *International Journal of STEM Education for Sustainability*.

Lee, G.-G., Mun, S., Shin, M.-K., & Zhai, X. (2025). Collaborative Learning with Artificial Intelligence Speakers. *Science & education*, *34*(2), 847–875. https://doi.org/10.1007/s11191-024-00526-y

Lee, G. G., & Zhai, X. (2024). Using ChatGPT for Science Learning: A Study on Pre-service Teachers' Lesson Planning. *Ieee Transactions on Learning Technologies*, *17*, 1643–1660. https://doi.org/10.1109/TLT.2024.3401457

Lee, U., Han, A., Lee, J., Lee, E., Kim, J., Kim, H., & Lim, C. (2024). Implication of a case study using generative AI in elementary school: Using Stable Diffusion for STEAM education. *The Journal of Applied Instructional Design*, *13*(2).

Mohana, M., Nandhini, K., & Subashini, P. (2022). Review on artificial intelligence and robots in STEAM education for early childhood development: The state-of-the-art tools and applications. In *Handbook of Research on Innovative Approaches to Early Childhood Development and School Readiness* (pp. 468–498).

Ngman-Wara, E. I. (2015). Ghanaian Junior High School Science Teachers' Knowledge of Contextualised Science Instruction. *Journal of Curriculum and Teaching*, *4*(1), 167–178.

Nikolovski, V., Trajanov, D., & Chorbev, I. (2025). Advancing AI in Higher Education: A Comparative Study of Large Language Model-Based Agents for Exam Question Generation, Improvement, and Evaluation. *Algorithms*, *18*(3), Article 144. https://doi.org/10.3390/a18030144

Nsengimana, T., Mugabo, L. R., Hiroaki, O., & Nkundabakura, P. (2020). Reflection on science competence-based curriculum implementation in Sub-Saharan African countries. *International Journal of Science Education*, 1–14.

Nyaaba, M., & Zhai, X. (2025). Developing a Theory-Grounded AI Tool for the Generation of Culturally Responsive Lesson Plans. *Computers and Education: Artificial Intelligence*, 100474. https://doi.org/https://doi.org/10.1016/j.caeai.2025.100474

Nyaaba, M., Zhai, X., & Faison, M. Z. (2024). Generative AI for Culturally Responsive Assessment in Science: A Conceptual Framework.

Ogunniyi, M. M. (2023). Culturally responsive science education for indigenous and ethnic minority students. In *Handbook of research on science education* (pp. 389–410). Routledge.

Price, J. F., & Grover, S. (2025). *Generative AI in STEM teaching: Opportunities and tradeoffs*. https://cadrek12.org/resources/generative-ai-stem-teaching-opportunities-and-tradeoffs

Roe, J., & Perkins, M. (2024). *Generative AI and agency in education: A critical scoping review and thematic analysis*. https://doi.org/10.48550/arXiv.2411.00631

Sakamoto, M., Tan, S., & Clivaz, S. (2024). Social, cultural and political perspectives of generative AI in teacher education: Lesson planning in Japanese teacher education.

*Exploring new horizons: Generative artificial intelligence and teacher education*, 178.

Shneiderman, B. (2022). *Human-Centered AI*. Oxford University Press. https://doi.org/10.1093/oso/9780192845290.001.0001

Şimşek, N. (2025). Integration of ChatGPT in mathematical story-focused 5E lesson planning: Teachers and pre-service teachers' interactions with ChatGPT. *Education and Information Technologies*, *30*(8), 11391–11462. https://doi.org/10.1007/s10639-024-13258-x

Thibeault-Orsi, G. (2022). *Contextualizing the Science, Technology, Engineering and Mathematics Gender Gap in European and sub-Saharan African Universities* University of Toronto (Canada)].

UNESCO. (2024). *AI Competency Framework for Teachers*. https://www.unesco.org/en/digital-education/g77-competencies

van den Berg, G., & du Plessis, E. (2023). ChatGPT and Generative AI: Possibilities for Its Contribution to Lesson Planning, Critical Thinking and Openness in Teacher Education. *Education Sciences*, *13*(10), 998.

Zhai, X. (2024). Transforming Teachers' Roles and Agencies in the Era of Generative AI: Perceptions, Acceptance, Knowledge, and Practices. *Journal of Science Education and Technology*. https://doi.org/10.1007/s10956-024-10174-0

Zhang, T., Ma, L., Yan, Y., Zhang, Y., Wang, K., Yang, Y., Guo, Z., Shao, W., You, Y., Qiao, Y., Luo, P., & Zhang, K. (2024). Rethinking Human Evaluation Protocol for Text-to-Video Models: Enhancing Reliability, Reproducibility, and Practicality. *ArXiv*, *abs/2406.08845*.

Zheng, Y. D., & Stewart, N. (2024). Improving EFL students' cultural awareness: Reframing moral dilemmatic stories with ChatGPT. *Computers and Education: Artificial Intelligence*, *6*, 100223.

Zhou, Z. Y. (2023). Evaluation of ChatGPT's Capabilities in Medical Report Generation [Article]. *Cureus Journal of Medical Science*, *15*(4), 14, Article e37589. https://doi.org/10.7759/cureus.37589